\documentclass[11pt,a4paper]{article}
\usepackage{jheppub}
\usepackage{graphicx, fancybox}
\usepackage{amsmath,amssymb}
\usepackage{subcaption}
\usepackage{setspace}
\usepackage{color}
\usepackage{soul}
\usepackage{url}
\DeclareMathOperator{\pf}{pf}
\DeclareMathOperator{\sgn}{sgn}
\DeclareMathOperator{\Tr}{tr}
\usepackage[normalem]{ulem}

\title{A new class of SYK-like models with maximal chaos}

\author[a]{Eric Marcus}
\author[a]{and Stefan Vandoren}
\affiliation[a]{Institute for Theoretical Physics, Utrecht University, Leuvenlaan 4, 3584 CE Utrecht, The Netherlands}
\emailAdd{e.j.marcus@uu.nl}
\emailAdd{S.J.G.Vandoren@uu.nl}

\abstract{We investigate a model closely related to both the original Sachdev-Ye-Kitaev (SYK) model and the $\mathcal{N}=1$ supersymmetric SYK model. It consists of $N$ real Majorana fermions and $M$ auxiliary bosons with Yukawa interactions. We consider the large $N$ and $M$ limit and keep the ratio $M/N$ fixed.  
The model has two branches characterized by the conformal dimensions of fields, which we compute as a function of the ratio $M/N$. One of the branches contains the supersymmetric saddle for $M=N$. Furthermore, we determine the Lyapunov exponent of the model and find maximal chaos independent of $M/N$. }


\arxivnumber{1808.01190}

\begin{document}
\maketitle
\flushbottom
\renewcommand{\figureautorefname}{figure}
\renewcommand{\appendixautorefname}{appendix}
\def\equationautorefname~#1\null{eq.\,(#1)\null}

\section{Introduction}
The Sachdev-Ye-Kitaev (SYK) model was introduced by Kitaev \citep{KitaevTalksSYK}, based on the original Sachdev-Ye model \citep{Sachdev:1992fk, Parcollet:1999, Georges:2000}.
One of the characterizing features of the model is the appearance of maximal chaos. This feature relates the model to black holes, which also show this behaviour \citep{Shenker:2013pqa, Shenker:2014cwa, Sachdev:2010um, Maldacena:2015waa, Sachdev:2015efa}. \\
In particular the SYK model is a (nearly) conformal field theory (CFT) in the infrared, and is assumed to have a nearly anti de sitter (AdS) dual in this regime \citep{Almheiri:2014cka, Engelsoy:2016xyb, Maldacena:2016upp}. For these low energies the model can be described by a Schwarzian, which also appears on the bulk side in the $AdS_2$ dilaton gravity.\\
The model has been intensively studied the past few years. There exists many generalizations including higher dimensions \citep{Berkooz:2016cvq, Gu:2016oyy, Jian:2017unn, Murugan:2017eto, Turiaci:2017zwd}, flavours \cite{Gross:2016kjj,Yoon:2017gut}, tunable chaos \cite{Chen:2017dbb}  and supersymmetry \cite{Fu:2016vas, Murugan:2017eto}.\\

In this paper we consider a particular model closely related to the $\mathcal{N}=1$ supersymmetric extension of SYK. Instead of having an equal number, $N$, of fermions and bosons we consider the case where we have $M$ bosons and $N$ fermions and study its behaviour as a function of the ratio $M/N$. \\
In \autoref{mdl} we will introduce the model and discuss in more detail the relation to the (supersymmetric) SYK model. Afterwards, in \autoref{efa}, we consider the effective action. We derive the equations of motion and consider the solutions at strong coupling. We find two families of solutions that we label by their conformal dimensions at $M=N$ (rational or irrational). Comparing the entropy of both solutions we determine that the rational solution is the dominant saddle for $M=N$. \\
In \autoref{chs} we compute the Lyapunov exponent and find that is independent of $M/N$ due to a subtle cancellation.

\section{The Model}\label{mdl}
The model consists out of $N$ Majorana fermions obeying $\{ \psi^i, \psi^j \} =\delta^{ij}$ and $M$ (auxiliary) bosons. We will use indices $a,b$ to denote the bosons and $i,j,k$ for the fermions (no ambiguity will arise). The Lagrangian is given as follows:
\begin{eqnarray}\label{startmn}
\mathcal{L}= \frac{1}{2} \sum\limits_{i=1}^N\psi^i \partial_{\tau} \psi^i -\frac{1}{2}\sum\limits_{a=1}^M \phi^a \, \phi^a +i \sum\limits_{a=1}^M \sum\limits_{i<j=1}^N C_{aij} \phi^a \, \psi^i \, \psi^j\ ,
\end{eqnarray} 
where $\psi$ denote the Majorana fermions and $\phi$ the bosons. 
The coupling $C_{aij}$ is defined to be antisymmetric in the last two indices, which are contracted with the Majorana fermions. In \citep{Bi:2017yvx} a similar term was studied as a perturbation upon ``normal" SYK. The fermions are dimensionless, whereas the bosons $\phi$ and couplings $C$ have dimension of $E^{1/2}$. \\ Notice that we have two parameters $M$ and $N$. We are interested in taking the limits of both $M$ and $N$ going to infinity but keeping $M/N$ fixed. In other words we have that $M=\alpha N$ for some fixed $\alpha$. From now on we will always assume that two $a$ indices are summed up to $M$ whilst the other $i,j,k,..$ are summed up to $N$. \\
Lastly, we let the coupling be disordered averaged by the following distribution:
\begin{eqnarray}\label{distrMN}
\langle C_{aij} \rangle &=&  0\ , \\
\langle C_{aij}^2 \rangle &=& \frac{2 J}{N^{3/2}M^{1/2}}\ .
\end{eqnarray}

\begin{figure}
\captionsetup{singlelinecheck = false, format= hang, justification=raggedright, labelsep=space}
   \centering
   \begin{subfigure}[b]{0.35\textwidth}
       \includegraphics[width=\textwidth]{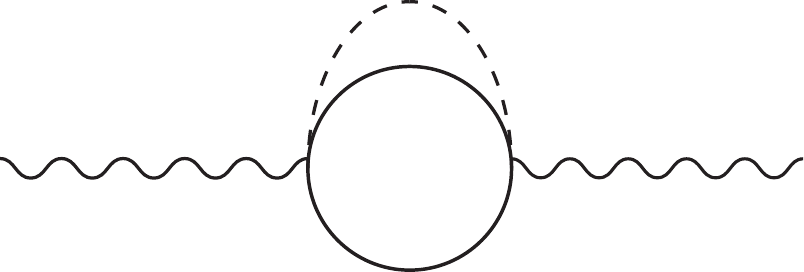}
       \caption{Proportional to $\sqrt{\frac{N}{M}}$.}
   \end{subfigure}\hspace{0.1\textwidth}
   \begin{subfigure}[b]{0.35\textwidth}
       \includegraphics[width=\textwidth]{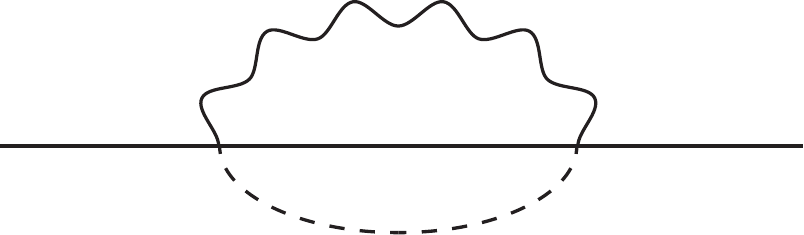}
       \caption{Proportional to $\sqrt{\frac{M}{N}}$.}
    \end{subfigure}
    \caption{In this figure we show the two one-loop corrections to the two point functions. The solid lines indicate fermions, the wiggly lines the bosons and the dotted line shows the disorder average. Below the diagrams we show the power of $M/N$ to which they are proportional.  }\label{twoptfig}
\end{figure}

Here $J$ has the dimension of energy and is larger than zero. We can now compute some basic one-loop diagrams for both the fermions and the bosons. We show the one-loop corrections to the two point functions in \autoref{twoptfig}, which are proportional to some power of $M/N$ (that can easily be checked). In fact one can check that any boson loop adds a factor of $\sqrt{\frac{M}{N}}$ and each fermion loop $\sqrt{\frac{N}{M}}$.\\

\subsection{Relation to SYK}
Let us first examine the relation to the original SYK model \cite{KitaevTalksSYK,Maldacena:2016hyu} with Hamiltonian
\begin{eqnarray}\label{orgsy}
H_{SYK} = -\frac{1}{4!} J_{ijkl} \, \psi^i \, \psi^j \, \psi^k \, \psi^l\ .
\end{eqnarray}

 To check the similarity we start by plugging in the algebraic equation of motion for $\phi^a$ back into the Lagrangian. The equation of motion is found to be $\phi^a =\frac{i}{2} C^a_{ij} \, \psi^i \, \psi^j$. After plugging it into \autoref{startmn} we obtain the Hamiltonian:
\begin{eqnarray}\label{testh}
H = \frac{1}{8} \, C_{aij} \, C_{akl} \, \psi^i \, \psi^j \, \psi^k \, \psi^l.
\end{eqnarray}

This is also the presentation that one can see in \citep{Bi:2017yvx}. We can then use the antisymmetry in the last two indices of $C_{aij}$ and the commutation relations of the Majorana fermions to rewrite this to:
\begin{eqnarray}
H =  \frac{1}{4!} \, \frac{1}{8} \,  C_{a[ij} \, C_{|a|kl]} \, \psi^i \, \psi^j \, \psi^k \, \psi^l +E_0\ ,
\end{eqnarray}
where we defined the constant $E_0 = -\frac{1}{16}C_{aij}^2$ (recall that $a$ is summed to $M$ and $i,j$ up to $N$). Comparing now to the standard SYK Hamiltonian, \autoref{orgsy}, we find:
\begin{eqnarray}
J_{ijkl} = -\sum\limits_{a=1}^M\frac{1}{8} C_{a[ij} \, C_{|a|kl]} \ .
\end{eqnarray}

The notation indicates that the asymmetry on the right hand side is only in $i,j,k$ and $l$, which in turn of course means that $J_{ijkl}$ is completely asymmetric. The above expression for the $J$ coupling shows us that the model is essentially obtained by performing a Hubbard-Stratonovich (HS) transformation on SYK. 
Of course apart from this HS transformation we have also chosen a different distribution (see \autoref{distrMN}) compared to SYK. This means that $J_{ijkl}$ are no longer the independent Gaussian variables and this is the cause of the differences between the models.

\subsection{Relation to supersymmetric SYK}
The $\mathcal{N}=1$ supersymmetric SYK model was introduced in \citep{Fu:2016vas}, the Lagrangian density is given by:
\begin{eqnarray}\label{susylag}
\mathcal{L} = \sum\limits_{i=1}^N \left(\frac{1}{2} \psi^i \partial_{\tau} \psi^i - \frac{1}{2} \phi^i \phi^i + \sum_{1\leq j<k \leq N}C_{ijk} \phi^i \psi^j \psi^k \, \right).
\end{eqnarray}
There are two important differences compared to the model described in \autoref{startmn}. The first important aspect is that there are $N$ bosons, which is the same as the number of fermions (which has to be true for supersymmetry). Secondly the coupling $C_{ijk}$ in the supersymmetric case has to be completely antisymmetric. Note that the equal number of bosons and fermions is also necessary for the antisymmetry in the coupling. \\

In other words, starting from \autoref{startmn} we can obtain the supersymmetric model by setting $M=N$ and making the coupling completely antisymmetric. It is precisely when the coupling is completely antisymmetric (and hence $M=N$) that the Lagrangian is invariant under supersymmetry transformations.

\section{Effective action and saddles}\label{efa}
To find the effective action we will follow the standard procedure of averaging over the disorder in $C_{aij}$ by using the replica trick (see appendices in \citep{Gross:2016kjj,Kitaev:2017awl}). As in the usual SYK case we will assume replica diagonal matrices. To justify this we have to compare $\overline{\log Z}$ and $\log \overline{Z}$ since assuming replica diagonal matrices corresponds to evaluating the latter instead of the former. The usual argument (see e.g. appendices in \citep{Kitaev:2017awl}) is to consider diagrams that are in $\log \overline{Z}$ but not in $\overline{\log Z}$. The leading diagram belonging to the former but not the latter is shown in \autoref{replfig} and as can be verified it is suppressed by $\frac{1}{N \, M}$. Thus in the large $N$ limit these contributions will be subdominant. Using replica symmetry, the result of the disorder average becomes:

\begin{figure}
\captionsetup{singlelinecheck = false, format= hang, justification=raggedright, labelsep=space}
\centering
\includegraphics[scale=0.82]{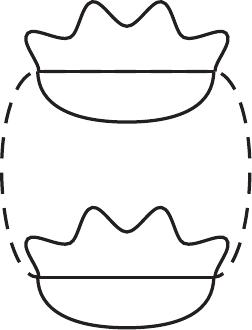}
\caption{This is the leading diagram that contributes to interactions between replicas. In the figure the top three lines would be associated with a different replica index than the other three below. One can check that this figure is proportional to $1/N$.}
\label{replfig}
\end{figure}

\begin{align}
S_{eff} = \frac{1}{2}&\int d\tau \left( \sum\limits_{i=1}^N \, \psi^i \, \partial_{\tau} \, \psi^i - \sum\limits_{a=1}^M \, \phi^a \, \phi^a \right) + \\
\notag &-\left( \sqrt{\frac{N}{M}}\frac{J}{2N^2} \int d\tau_1 d\tau_2 \sum\limits_{a,(j,k)} \left( \phi^a (\tau_1) \phi^a (\tau_2) \right) \,\left( \psi^j (\tau_1) \psi^j (\tau_2) \right) \, \left( \psi^k (\tau_1) \psi^k (\tau_2) \right)\right).
\end{align}

For the last term we introduced brackets below the sum to indicate that $j,k$ sum up to $N$ whilst $a$ sums up to $M$.
We now introduce bilocal fields for both the fermions and bosons as follows:
\begin{eqnarray}\label{start1}
&\delta\left(G_{\psi }(\tau_1,\tau_2) - \frac{1}{N} \sum\limits_{i=1}^N  \psi^i (\tau_1) \, \psi^i (\tau_2) \right) = \\
&\notag \int d\Sigma_{ \psi}(\tau_1,\tau_2) \, \exp \left \{ -\frac{N}{2} \, \Sigma_{\psi }(\tau_1,\tau_2) \left( G_{ \psi}(\tau_1,\tau_2) - \frac{1}{N} \sum\limits_{i=1}^N  \psi^i (\tau_1) \, \psi^i (\tau_2) \right)    \right \},
\end{eqnarray}
\begin{eqnarray}
&\delta\left(G_{\phi}(\tau_1,\tau_2) - \frac{1}{M} \sum\limits_{a=1}^M  \phi^a (\tau_1) \, \phi^a (\tau_2) \right) = \\
&\notag \int d\Sigma_{ \phi}(\tau_1,\tau_2) \, \exp \left \{ -\frac{M}{2} \, \Sigma_{\phi }(\tau_1,\tau_2) \left( G_{\phi}(\tau_1,\tau_2) - \frac{1}{M} \sum\limits_{i=1}^M  \phi^i (\tau_1) \, \phi^i (\tau_2) \right)    \right \}.
\end{eqnarray} 

We insert them into the partition function by Lagrange multipliers. Afterwards we are only left with Gaussian integrals for both the fermions and bosons. Completing these leads to:
\begin{eqnarray}\label{efac}
\frac{S_{eff}}{N} = -\log \pf \left( \partial_{\tau}- \Sigma_{\psi }(\tau) \right)+ \frac{M}{2\,N} \log \det \left(-1-\Sigma_{ \phi}(\tau) \right) \\
\notag + \frac{1}{2} \int d\tau_1 d\tau_2 \left[ \Sigma_{ \psi}(\tau_1,\tau_2) \, G_{ \psi}(\tau_1,\tau_2) + \frac{M}{N}\Sigma_{ \phi}(\tau_1,\tau_2) \, G_{ \phi}(\tau_1,\tau_2) \right. \\
\notag \left.  - J \, \sqrt{\frac{M}{N}} \, G_{ \phi}(\tau_1,\tau_2) \, G^2_{ \psi}(\tau_1,\tau_2)  \right].
\end{eqnarray}

Where on the left hand side we divided out a factor of $N$, but could just as well have taken out $M$ (recall that $M/N$ is fixed).\\
Let us now vary with respect to $G_{\phi}$ and $G_{ \psi}$ to obtain the self energies:
\begin{eqnarray}\label{selfenergymn}
\Sigma_{\psi} &=& 2 J \sqrt{\frac{M}{N}} \, G_{\phi} \, G_{\psi}\ , \\
\notag \Sigma_{\phi} &=& J \sqrt{\frac{N}{M}} \, G_{\psi}^2\ .
\end{eqnarray}
These equations can also be obtained using the melonic structure of the Feynman diagrams at large $N$ and $M$, just as in ordinary SYK.
The Schwinger-Dyson equations are obtained by varying with respect to the $\Sigma$ (we assume time translation symmetry and go to Fourier space):
\begin{eqnarray}\label{sdmn}
G_{\psi}^{-1}(i\omega) &=& -i\omega - \Sigma_{\psi}(i\omega)\ , \\
\notag G_{\phi}^{-1}(i\omega) &=& -1-\Sigma _{\phi}(i\omega)\ .
\end{eqnarray}

\subsection{Two saddle points}
In order to solve the above equations we have to assume the strong coupling limit $\beta J \gg 1$. This implies that in \autoref{sdmn} we can ignore the first terms on the right hand side. Hence we can write the equations as follows (we have Fourier transformed back to time):
\begin{eqnarray}\label{confeqnmn}
\notag&\int d\tau' \, G_{\psi}(\tau,\tau') \, \Sigma_{\psi}(\tau',\tau'') = 2J \sqrt{\frac{M}{N}}\int d\tau'G_{\psi}(\tau,\tau') \, G_{\phi}(\tau',\tau'')G_{\psi}(\tau',\tau'') = \\
& =-\delta(\tau-\tau'')\ ,
\end{eqnarray}
\begin{eqnarray}\label{confeqn2mn}
\notag&\int d\tau' \, G_{\phi}(\tau,\tau') \, \Sigma_{\phi}(\tau',\tau'') = J \sqrt{\frac{N}{M}}\int d\tau'G_{\phi}(\tau,\tau') \,G^2_{\psi}(\tau',\tau'') = \\
& =-\delta(\tau-\tau'')\ .
\end{eqnarray}
We then use the following (conformal) form for the two point functions:
\begin{eqnarray}\label{ansz}
G_{\psi}(\tau) &=& A \, \frac{\sgn(\tau)}{|\tau|^{2 \Delta_{\psi}}}\ ,\\
G_{\phi}(\tau) &=& B \, \frac{1}{|\tau|^{2 \Delta_{\phi}}}\ .
\end{eqnarray}
To obtain conditions on the conformal dimensions we plug these into the saddle point equations above, \autoref{confeqnmn} and \autoref{confeqn2mn}. Afterwards we Fourier transform using \citep{Fu:2016vas, Maldacena:2016hyu}:
\begin{eqnarray}
\int d\tau e^{i \, \omega \, \tau} \, \frac{\sgn(\tau)}{| \tau |^{2\Delta}} &=& 2 i \, \cos(\pi \Delta) \, \Gamma (1-2\Delta) \, \sgn( \omega ) \, |\omega |^{2\Delta - 1}, \\
\int d\tau e^{i \, \omega \, \tau} \, \frac{1}{| \tau |^{2\Delta}} &=& 2 \, \sin(\pi \Delta) \, \Gamma (1-2\Delta) \,  |\omega |^{2\Delta - 1} .
\end{eqnarray}
Some other useful relations for $\Gamma$ functions are
\begin{eqnarray}\label{gamid}
\Gamma (1-2\Delta) &=& \frac{2^{-2\Delta}\, \sqrt{\pi}}{\cos (\pi \Delta )} \, \frac{\Gamma (1-\Delta)}{\Gamma \left(\frac{1}{2} + \Delta \right)}\ , \\
\frac{\Gamma (1- \Delta) \, \Gamma(\Delta)}{\Gamma \left(\frac{1}{2}+\Delta\right) \, \Gamma \left(\frac{3}{2} - \Delta \right)} &=& \frac{2}{1-2\Delta} \, \frac{\cos (\pi \Delta)}{\sin (\pi \Delta)}\ .
\end{eqnarray}
After plugging this al in we obtain the following relations:
\begin{eqnarray}\label{coneq}
A^2 B \sqrt{\frac{M}{N}}\,\frac{4\, \pi \, J}{1-2 \Delta_{\psi}}\,\frac{\cos(\pi \Delta_{\psi})}{\sin(\pi \Delta_{\psi})} \, | \omega |^{2(2\Delta_{\psi}+\Delta_{\phi})-2} = 1\ , \\
\notag A^2 B \sqrt{\frac{N}{M}}\,\frac{2\, \pi \, J}{1-4 \Delta_{\psi}}\, \tan(2\pi \Delta_{\psi}) | \omega |^{2(2\Delta_{\psi}+\Delta_{\phi})-2} = 1\ . 
\end{eqnarray}
These relations (for $M=N$) have also been derived in \cite{Fu:2016vas}.
By comparing the frequency dependent parts we obtain the first condition on the conformal dimensions:
\begin{eqnarray}\label{constr1mn}
2\Delta_{\psi}+\Delta_{\phi} = 1\ .
\end{eqnarray}
As a side note, under this condition the saddle point equations have the conformal symmetry, very analogous to the original SYK model:
\begin{eqnarray}\label{confsym}
G_{\psi}(\tau,\tau') &=& |f'(\tau)\, f(\tau')|^{\Delta_{\psi}} \, G_{\psi}\left(f(\tau),f(\tau')\right) \ , \\
\notag G_{\phi}(\tau,\tau') &=& |f'(\tau)\, f(\tau')|^{\Delta_{\phi}} \, G_{\phi}\left(f(\tau),f(\tau')\right) \ .
\end{eqnarray}

Where $f(\tau)$ a smooth function (in one dimension $\text{Conf}(\mathbb{R}) \cong \text{Diff}(\mathbb{R})$). To obtain results for finite temperature we use this symmetry with $f$ being the exponential map for example. \\

Coming back to \autoref{coneq}, we can obtain another constraint by taking the quotient, which yields the (transcendental) equation:
\begin{eqnarray}\label{soleqnnmn}
\frac{N}{M} \,\tan (\pi  \Delta_{\psi} ) \, \tan (2 \pi  \Delta_{\psi}) = \frac{2(1-4 \, \Delta_{\psi})}{1-2 \Delta_{\psi}}\ .
\end{eqnarray}

This result, for $M=N$, is also obtained in \citep{Fu:2016vas}, although it contains some typos. In \citep{Bi:2017yvx} it is also shown for $M \neq N$, albeit in a different form.  \\
The second condition, \autoref{soleqnnmn}, can also be recast to an equation for $\Delta_{\phi}$ using \autoref{constr1mn}:
\begin{eqnarray}\label{soleqnn2mn}
-4+\frac{2}{\Delta_{\phi}}-\frac{N}{M}\, \frac{\tan(\pi \Delta_{\phi})}{\tan (\frac{1}{2} \pi \Delta_{\phi})}=0\ .
\end{eqnarray}

\subsubsection{The case $M=N$}
First we solve \autoref{soleqnnmn} for $M$ being equal to $N$. This case overlaps with supersymmetric SYK (as commented upon in the introduction) and we find the same solutions as in \citep{Fu:2016vas}. The first solution is given by:
\begin{eqnarray}\label{soln1}
\notag \Delta_{\psi} &=& \frac{1}{6}\ , \\
\Delta_{\phi} &=& \frac{2}{3}\ , \\
\notag A^2 B &=& \frac{1}{6 \, \pi \, J \, \sqrt{3}}\ .
\end{eqnarray}

We label this solution as the ``rational" solution. In the supersymmetric model this solution is the one that preserves supersymmetry. In that case, the supersymmetric Ward identity $G_{\phi} = \partial_{\tau}G_{\psi}$, together with \autoref{ansz} implies $\Delta_{\phi} = \Delta_{\psi}+\frac{1}{2}$ \citep{Fu:2016vas}, obviously obeyed by \autoref{soln1}. \\
There is another solution with positive conformal dimensions, it is however irrational:
\begin{eqnarray}\label{soln2}
\notag \Delta_{\psi} &=& 0.350585...\ , \\
\Delta_{\phi} &=& 0.29883... \ ,\\
\notag A^2 B &=& \frac{0.589161...}{4 \pi J}\ .
\end{eqnarray}

As one can easily check this does not satisfy $\Delta_{\phi} = \Delta_{\psi}+\frac{1}{2}$ and hence would break supersymmetry. A similar situation arises in \citep{Anninos:2016szt} where there are also two solutions, one preserving and one breaking the supersymmetry.

\subsubsection{Arbitrary $M$ and $N$}
\begin{figure}
\captionsetup{singlelinecheck = false, format= hang, justification=raggedright,  labelsep=space}
\includegraphics[scale=0.76]{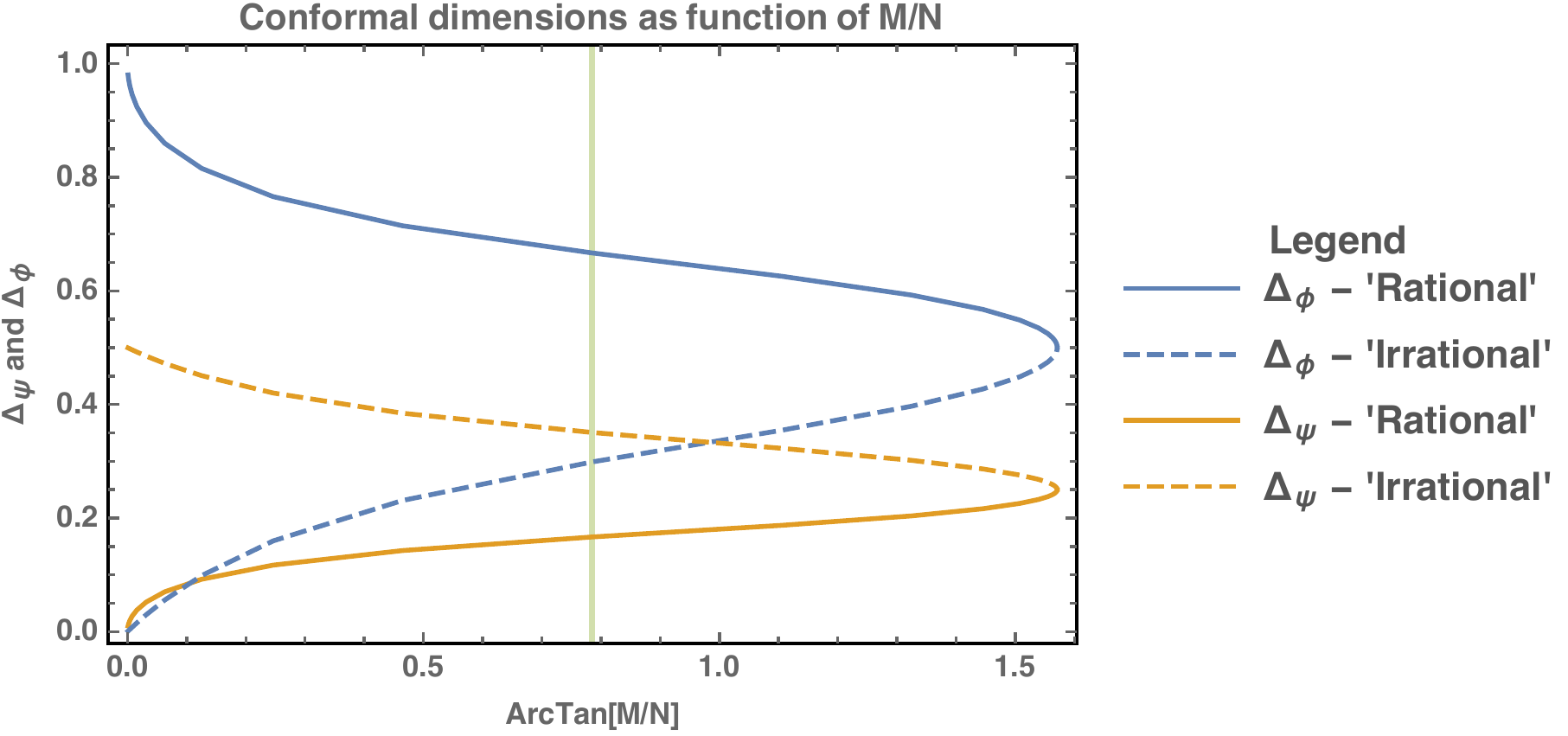}
\caption{This figure shows the conformal dimensions as a function of $M/N$. The two solutions are labelled by their (ir)rational behaviour at $M=N$ (which may differ for other values), see \autoref{soln1} and \autoref{soln2}. The green line represents this point where $M/N=1$. }
\label{confdimfig}
\end{figure}

Let us now vary the ratio $M/N$ and find the conformal dimensions as a function of this ratio. We solve \autoref{soleqnnmn} numerically and show the results in \autoref{confdimfig}. There are two ``families" of solutions, labelled by their behaviour at $M=N$. The rational solution was also found in \citep{Bi:2017yvx}. 

When $M/N$ becomes large the rational and irrational flow to the same point. This can be understood by considering the defining equations \autoref{soleqnnmn} and \autoref{soleqnn2mn}. When one takes the limit of $M/N$ going to infinity there is only one solution left:
\begin{eqnarray}
&\Delta_{\psi} = \frac{1}{4}\ , \\
&\notag \Delta_{\phi} = \frac{1}{2}\ .
\end{eqnarray}

Similarly, the behaviour for small $M/N$ can be understood by taking the appropriate limits in the defining equations. This is equivalent to considering the limit $N/M$ going to infinity in \autoref{soleqnn2mn}. Consider the following two limits:
\begin{eqnarray}
\lim\limits_{\Delta_{\phi}\rightarrow \; 1} \; \;  \frac{\tan\left(\pi \Delta_{\phi}\right)}{\tan\left(\frac{\pi}{2}\Delta_{\phi}\right)} = 0 \; , \qquad \lim\limits_{\Delta_{\phi}\rightarrow \; 0} \; \;  \frac{\tan\left(\pi \Delta_{\phi}\right)}{\tan\left(\frac{\pi}{2}\Delta_{\phi}\right)} = 2\ .
\end{eqnarray}

Applying these in \autoref{soleqnn2mn} shows us that for small $M/N$ we either need to consider the case where $\Delta_{\phi}$ is very small or the case where it goes to one. The latter corresponds to the rational family. For the irrational $\Delta_{\phi}\ll1$ case we find from \autoref{soleqnn2mn} that it behaves exactly as:
\begin{eqnarray}
\Delta_{\phi} = \frac{M}{N}\ .
\end{eqnarray}

 The corresponding dimensions for the fermion can of course be found by \autoref{constr1mn}. 
Lastly let us investigate the rational $\Delta_{\psi}$ for small $M/N$. Observing \autoref{confdimfig}, we assume that $\Delta_{\psi}$ is small and consider \autoref{coneq}. We can then solve exactly as follows:
\begin{eqnarray}\label{exact}
A^2B = \frac{1}{4 \,\pi J}\ ,\qquad \Delta_{\psi} = \frac{1}{\pi} \, \sqrt{\frac{M}{N}}\ .
\end{eqnarray}

\subsection{Dominant saddle}
In this section we will determine what is the dominant saddle by comparing the entropies of both solutions. In particular we consider the case $M/N=1$, since we know here exactly the behaviour of the rational solution as a function of $q$ (see below). 
For the computation we will follow \citep{Fu:2016vas} and use the model for a $q$-interaction (meaning a vertex with one boson and $q-1$ fermions, with $q$ odd), see \autoref{qapp} for an overview of the changes. The free energy becomes:
\begin{eqnarray}\label{efacq}
\frac{\log(Z)}{N} = -\log \pf \left( \partial_{\tau}- \Sigma_{\psi \psi}(\tau) \right)+ \frac{M}{2N} \log \det \left(-1-\Sigma_{\phi \phi}(\tau) \right) \\
\notag + \frac{1}{2} \int d\tau_1 d\tau_2 \left[ \Sigma_{\psi \psi}(\tau_1,\tau_2) \, G_{\psi \psi}(\tau_1,\tau_2) + \frac{M}{N}\Sigma_{\phi \phi}(\tau_1,\tau_2) \, G_{\phi \phi}(\tau_1,\tau_2) \right. \\
\notag \left. - J \, \sqrt{\frac{M}{N}} \, G_{\phi \phi}(\tau_1,\tau_2) \, G^{q-1}_{\psi \psi}(\tau_1,\tau_2)  \right].
\end{eqnarray}

Now we derive with respect to $q$ (we continue the values of $q$ to the reals) such that we don't have to evaluate the first terms. We take the fields to be on-shell such that we only need to explicitly take the partial derivative of the last term
\begin{eqnarray}
\partial_q \frac{\log(Z)}{N} =\frac{J}{2} \, \sqrt{\frac{M}{N}} \int d\tau \, d\tau' G_{\phi}(\tau-\tau') \, \log\left(G_{\psi}(\tau-\tau')\right)\,G^{q-1}_{\psi}(\tau-\tau')\ ,
\end{eqnarray}
where the $G$s are now the finite temperature versions, obtained by conformal symmetry mentioned before (\autoref{confsym}):
\begin{eqnarray}\label{fint}
G_{\psi} (\tau) = A \left(\frac{\pi}{\beta \, \sin \left(\frac{\pi \tau}{\beta}\right)} \right)^{2\Delta_{\psi}} \, \sgn (\tau) \ , \qquad G_{\phi} (\tau) = B \left(\frac{\pi}{\beta \, \sin \left(\frac{\pi \tau}{\beta}\right)} \right)^{2\Delta_{\phi}} \ .
\end{eqnarray}

The integral can then be computed straightforwardly (using the periodicity in the $\tau$ variables):
\begin{eqnarray}\label{int11}
\partial_q \frac{\log(Z)}{N} =\sqrt{\frac{M}{N}} \, \frac{J}{2}A^{q-1}\,B \, \pi^2 \left[2 \Delta_{\psi} + \beta \, C\right].
\end{eqnarray}

Where $C$ is a constant independent of $\beta$.
The constant term is a diverging quantity independent of $q$ contributing to the ground state energy but will not contribute to the entropy, similar to the scenario in \citep{Fu:2016vas}. \\
It is important to note that apart from the overall factor, the $M/N$ dependence is also in $A^{q-1}B$ (\autoref{genqconst}) and the conformal dimension $\Delta_{\psi}$ (\autoref{confdimfig}). For now we will consider the case $M=N$. \\

The entropies $S_R$ and $S_I$ are labelled by their rational of irrational origin, see \autoref{soln1} and \autoref{soln2} respectively, and given by:
\begin{eqnarray}\label{csts2}
S_R &:=&  \int r(q) \, dq = R(q) + C_R\ ,\\
\notag S_I &:=& \int i(q) \, dq = I(q) + C_I\ ,
\end{eqnarray}
where $C_R$ and $C_I$ are integration constants. Furthermore we have called the $q$ dependent parts $R(q)$ and $I(q)$ (these do not contain constants).
\subsubsection{The rational branch}
Let us first consider the rational branch. In this case we can always solve the exact dependence of the conformal dimension on $q$ (see \autoref{qapp}):
\begin{eqnarray}
\Delta_{\psi}^R = \frac{1}{2q}\ .
\end{eqnarray}
The above expression means that the integrand in \autoref{csts2}, $r(q)$, can be computed using \autoref{int11}:
\begin{eqnarray}
r(q) = \frac{\pi}{4 q^2} \tan \left(\frac{\pi}{2q}\right)\ ,
\end{eqnarray}
and hence we can compute the entropy for the rational case:
\begin{eqnarray}\label{ent1}
\frac{S_R}{N} = \frac{1}{2} \log \left(\cos\left(\frac{\pi}{2 q}\right)\right)+C_R \ .
\end{eqnarray}

To fix the integration constant we will consider the limit $q \rightarrow \infty$. For the case $M/N=1$ we can follow exactly \citep{Fu:2016vas}, section II.C. There the results in a large $q$ expansion are obtained:
\begin{align}
&G_{\psi} = \frac{1}{2} \sgn (\tau) + \frac{1}{2q} g_{\psi}(\tau) \ , \qquad G_{\phi} = -\delta(\tau) + \frac{1}{2q}\, g_{\phi}(\tau) \ , \\
&\frac{\log Z}{N} = \frac{1}{2} \log 2 + \frac{1}{4q^2} \left( - \frac{v^2}{4} +v \, \tan \frac{v}{2} \right) \ .
\end{align}

Where $v$ is an integration constant related to $\beta J$ \citep{Fu:2016vas}. The expansion in large $q$ can still be made in a similar manner (although the functions $g_x (\tau)$ may now contain factors of $M/N$).\\
From these expressions above it becomes clear that the $q\rightarrow \infty$ limit reduces to free fermions. It also allows us to fix the constant $C_R$ since:
\begin{eqnarray}
\lim\limits_{q\rightarrow\infty} &\;\frac{1}{2} \log \left(\cos\left(\frac{\pi}{2 q}\right)\right)=0 \ , \\
\lim\limits_{q\rightarrow\infty} &\; \frac{S_R}{N} = C_R = \frac{1}{2} \log 2 \ .
\end{eqnarray}

Where in the last line we used the above observation that it should reduce to a free fermion entropy.

\subsubsection{The irrational branch}
\begin{figure}
\captionsetup{singlelinecheck = false, format= hang, justification=raggedright, labelsep=space}
\includegraphics[scale=0.76]{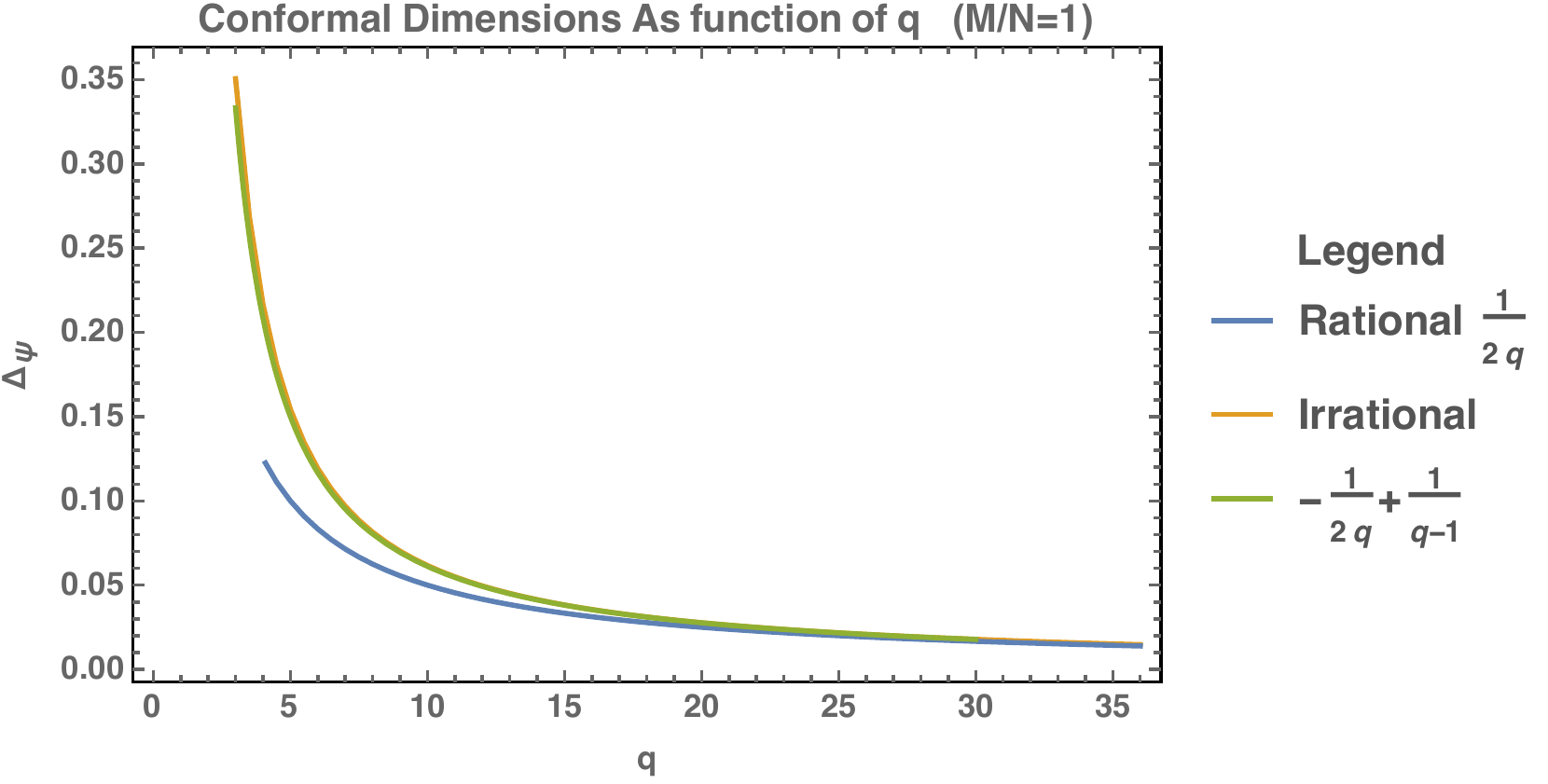}
\caption{Here we plot the dependence of the conformal dimensions on the number of interactions $q$. Although we do not have the exact results for the irrational case, it is well approximated by the guessed solution in green. it can be seen that for large enough $q$ the solutions approach one another.}
\label{funcq}
\end{figure}

Unfortunately we can't solve analytically the $q$ dependence of the irrational solution. We did manage to find a good fit by $-\frac{1}{2q}+\frac{1}{q-1}$, which matches the numerical results very well for large $q$. In fact, we only find small deviations for low values of $q$. In \autoref{funcq} we plot the numerical results, the best-fit solution and the rational solution. \\

\begin{figure}
\captionsetup{singlelinecheck = false, format= hang, justification=raggedright,  labelsep=space}
\includegraphics[scale=0.76]{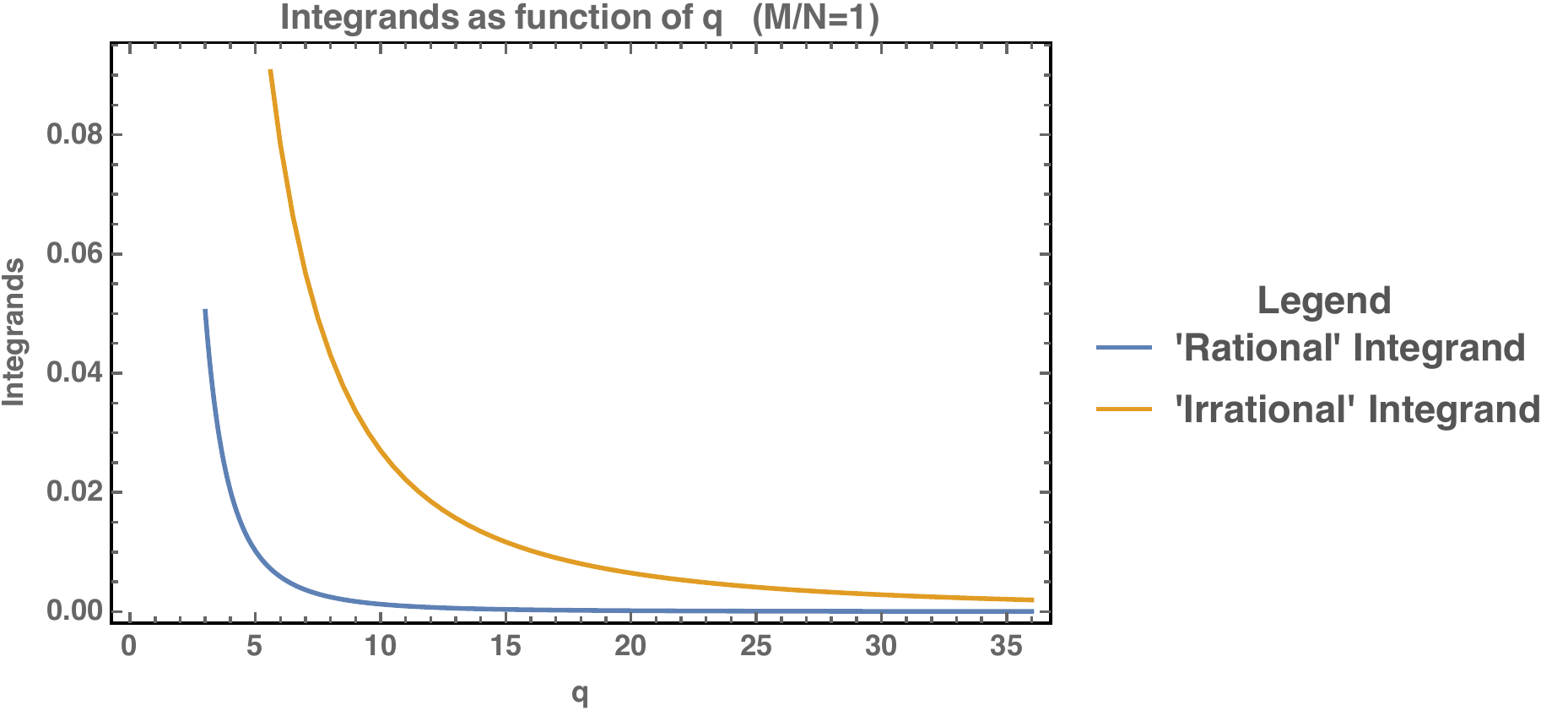}
\caption{This plot shows the dependence of the integrands in \autoref{csts2} on the number of interactions $q$. We can see that the irrational integrand ($i(q)$) is larger than the rational one ($r(q)$).}
\label{intfig}
\end{figure}

To conclude which of the entropies is bigger (i.e. which is the dominant saddle) we will investigate the integrands as a function of $q$, see \autoref{intfig}. From this plot we can see that the irrational integrand is bigger than the rational one and as $q$ increases their difference decreases.

We now investigate again the behaviour for large $q$. By using the approximate solution (see \autoref{funcq}) we can obtain an expression for $i(q)$ at large $q$. This is done by using the approximate solution in \autoref{int11}:
\begin{eqnarray}
i(q) \overset{q\gg 1}{=} \frac{\pi  (q+1) ((q-2) q-1) \tan \left(\frac{\pi  (q+1)}{2 (q-1) q}\right)}{4 (q-1)^3 q^2} \approx \frac{\pi ^2}{8 q^3} + \frac{\pi ^2}{2 q^4} + \mathcal{O} \left(\frac{1}{q^5}\right).
\end{eqnarray}

From this we can guess the leading order behaviour of $I(q)$ in \autoref{csts2}:
\begin{eqnarray}\label{largeqbeh}
I(q) &\overset{q\gg 1}{\approx}& -\frac{\pi^2}{16 q^2}-\frac{\pi^2}{6q^3}+\mathcal{O} \left(\frac{1}{q^4}\right) \ , \\
R(q)  &\overset{q\gg 1}{\approx}&-\frac{\pi ^2}{16 q^2} -\frac{\pi ^4}{384 q^4}+ \mathcal{O}  \left(\frac{1}{q^5}\right) \ .
\end{eqnarray}

Where we also gave the rational behaviour for large $q$, which can be obtained from \autoref{ent1}. It can be seen that for these large $q$ we have $I(q)<R(q)$. Now to fix the integration constant for the irrational case (see \autoref{csts2}) we can take the strict $q \rightarrow \infty$ limit. In this limit the two solutions (irrational and rational) will coincide since their conformal dimensions will equal. Hence we also have that their entropies must coincide:
\begin{eqnarray}\label{Iprop}
0=\lim\limits_{q\rightarrow \infty}\left(S_I - S_R \right) = \lim\limits_{q\rightarrow \infty} I(q) + C_I - C_R \ .
\end{eqnarray}

Using \autoref{largeqbeh} it becomes clear that the first term is zero and hence $C_I =C_R=\frac{1}{2}\log2$.
We can then conclude that for large $q$: $S_I<S_R$. Further more since the slope of $R(q)$ is always smaller than that of $I(q)$ (since $i(q)>r(q)$) we find that this conclusion holds for any $q$. \\

To conclude this analysis, we found that the rational solution is the dominant saddle in this model at $M=N$. To extend the analysis for arbitrary $M/N$ we would need to find a best-fit of the conformal dimensions as a function of $q$ similar to above. Afterwards we can follow the same procedure to figure out the dominant one. Since this numerical analysis is quite tedious, we have left it for future work.

\section{Chaos}\label{chs}
In this section we will investigate the chaos or Lyapunov exponent of the model as a function of the ratio $M/N$. We will first review shortly the basics of such a computation and then move on to our model. The main tool for quantifying quantum chaos are so called Out of Time Order Correlators (OTOC) \cite{1969JETP...28.1200L,Almheiri:2013hfa,Shenker:2013pqa,KitaevTalksHawking,Roberts:2014ifa}.  For a more elaborate review of chaos and calculating these correlators see chapter 8 in \cite{Murugan:2017eto}, the first section of \citep{Maldacena:2015waa} and a discussion in \cite{Kitaev:2017awl}.\\

From a quantum mechanical point of view we can take two arbitrary Hermitian operators $V$ and $W$ and consider the commutator $[W(it),V(0)]$ (with real time $t \in \mathbb{R}$). The argument of the operator is imaginary since we consider it to be Euclidean time, as will be the case for our operators later on. The commutator describes the influence of small changes of $V$ on later measurements of $W$ (or the other way around). One particular indicator of these effects of chaos, which we will also use, puts the operators on the thermal circle \citep{Maldacena:2015waa}:
\begin{eqnarray}\label{comu}
\langle \;\left[V(0),W(it)\right] \, \left[V \left(\beta / 2 \right), W \left(\beta / 2+it \right) \right] \; \rangle \ .
\end{eqnarray}

Where the brackets $\langle \rangle$ denote the thermal trace, the precise factors of $\beta$ will not be important for us. For late enough times $t$ (to be precise, between the dissipation and scrambling time \cite{Maldacena:2015waa}), quantum chaos dictates that this correlator will grow exponentially. By considering all the terms that arise in the above correlator one can show \cite{Maldacena:2015waa} that the exponential growth of the correlator arises due to the exponential behaviour of the related correlator:
\begin{eqnarray}\label{otoc}
F(t) = \langle \, V(0) \, W(\beta/4 +it) \, V(\beta / 2)\, W(3 \beta / 4 + it) \, \rangle \ .
\end{eqnarray}

These out of time order correlators $F(t)$ are usually studied in the context of quantum chaos, and we will use these as well. Schematically the OTOC \autoref{otoc} behaves as \cite{KitaevTalksHawking,Maldacena:2015waa,Murugan:2017eto}:
\begin{eqnarray}
F(t) = 1 - \frac{1}{N}e^{\lambda_L t}+ \dots 
\end{eqnarray}

The exponent $\lambda_L$ is called the Lyapunov exponent and it quantifies the chaos of the system. In the coming section our goal is to extract this Lyapunov exponent from the OTOCs. In general one can follow two approaches. The most obvious one is to compute the full four point function and continue these Euclidean correlators to real time. An easier option, however, is to consider the so called retarded kernel and its eigenfunctions \cite{KitaevTalksHawking,KitaevTalksSYK,Maldacena:2016hyu}. In the context of ladder diagrams, kernels are the operators that add one more ladder to the diagram. For the OTOCs it has to be the retarded kernel due to the complex time contours specified by OTOCs similar to \autoref{comu}.  For a review of this procedure including the complex time contours, ladder diagrams and the application to ordinary SYK see \cite{Murugan:2017eto}.\\

The key idea of this procedure is to consider an exponentially growing OTOC on which the kernel(s) are acting. Under the assumption of this exponential growth one can find that it is precisely the eigenfunctions of the kernel with eigenvalue one that govern the chaotic behaviour. More intuitively, the growth rate of OTOC is determined by the demand that adding another ladder should not change the total sum. In the rest of this section we explain this procedure in more detail.

\subsection{Retarded kernels}
Let us now turn to our model and consider the four point functions (or OTOCs) that we want to compute. In \citep{Peng:2017kro,Peng:2017spg} the chaos is also calculated for similar circumstances and we will comment upon this method at the end. \\
We will consider the four point functions $\langle \, \psi \, \psi \, \psi \, \psi \, \rangle$, $\langle \, \psi \, \psi \, \phi \, \phi \, \rangle$, $\langle \, \phi \, \phi \, \psi \, \psi \, \rangle$ and $\langle \, \phi \, \phi \, \phi \, \phi \, \rangle$. This is because acting with kernels on these diagrams will result in mixing between them and hence we can not consider them separately. The explicit OTOCs we will consider are of the form:
\begin{eqnarray}\label{otocour}
F_{\psi \psi}(t_1,t_2) = \Tr \left[ y \, \psi (t_1) \, y \, \psi(0) \, y \, \psi(t_2) \, y \, \psi (0)\right] \ ,
\end{eqnarray}
Where $y$ is defined as $y^4 = \rho (\beta)$. Diagrammatically these OTOCs are four point functions (ladder diagrams) with an arbitrary large amount of rungs.
The other combinations of $\psi$ and $\phi$ listed above have similar expressions and are denoted by $F_{\psi \phi}$, $F_{\phi \psi}$ and $F_{\phi \phi}$. The two subscripts of $F_{ij}$ denote the two incoming and two outgoing species, respectively.  \\

 \begin{figure}
\captionsetup{singlelinecheck = false, format= hang, justification=raggedright, labelsep=space}
   \centering
   \begin{subfigure}[b]{0.23\textwidth}
       \includegraphics[width=\textwidth]{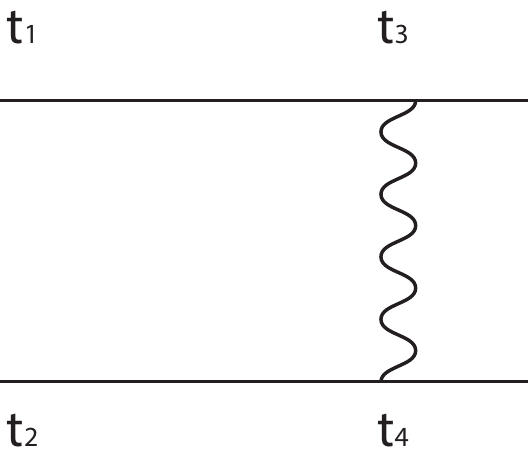}
       \caption{$K_{11}$ kernel}
   \end{subfigure}\hspace{0.08\textwidth}
   \begin{subfigure}[b]{0.25\textwidth}
       \includegraphics[width=\textwidth]{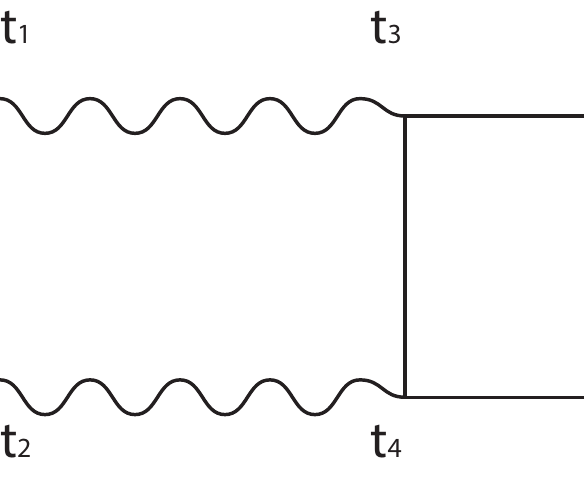}
       \caption{$K_{21}$ kernel}
    \end{subfigure}\hspace{0.08\textwidth}%
      \begin{subfigure}[b]{0.25\textwidth}
       \includegraphics[width=\textwidth]{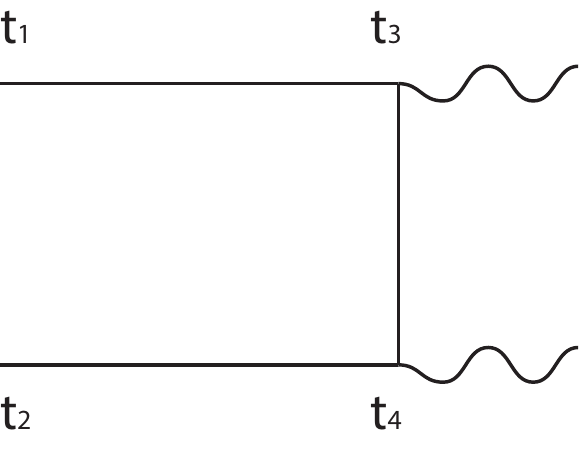}
       \caption{$K_{12}$ kernel}
    \end{subfigure}
    \caption{Here we show the relevant kernels for the chaos computation. The subscripts of the kernels denote that they are elements of a matrix. The total matrix acts on a vector consisting of diagrams which starts with either two $\psi$ or two $\phi$ lines.}\label{kernelfigs}
\end{figure}

Let us now consider all the (retarded) kernels necessary for our model, which we draw in \autoref{kernelfigs}. Note that there is no kernel $K_{22}$ since there is no such interaction in the Lagrangian. It then becomes clear that acting with the $K_{12}$  and $K_{21}$ kernels causes mixing between the four point functions. \\
To get expressions for them we need first the necessary propagators in the diagrams. For the horizontal propagators we need the retarded ones (due to the complex time contours, see \cite{Murugan:2017eto}):
\begin{eqnarray}\label{ret}
G^{\psi}_R (t) = \left(\langle \psi^i (it) \, \psi^i (0)\rangle + \langle \psi^i(0) \, \psi^i(it) \rangle \right) \theta(t)\ , \\
\notag G^{\phi}_R (t) = \left(\langle \phi^a (it) \, \phi^a (0)\rangle - \langle \phi^a(0) \, \phi^a(it) \rangle \right) \theta(t)\ .
\end{eqnarray}

Recall that the arguments are imaginary since we consider complex Euclidean time. We can then use the finite temperature two point functions from \autoref{fint} to find:
\begin{eqnarray}
G^{\psi}_R (t) =   \frac{2 \,A \,\cos(\pi \Delta _{\psi})\, \pi^{2\Delta_{\psi}}}{\left(\beta \sinh\left(\frac{\pi t}{\beta}\right)\right)^{2\Delta_{\psi}}} \theta(t)\ .
\end{eqnarray}
And similarly for $\phi$:
\begin{eqnarray}
G^{\phi}_R (t) = - \frac{2i \,B \,\sin(\pi \Delta _{\phi})\, \pi^{2\Delta_{\phi}}}{\left(\beta \sinh\left(\frac{\pi t}{\beta}\right)\right)^{2\Delta_{\phi}}} \theta(t)\ .
\end{eqnarray}

Lastly we need the ladder rung ($lr$) propagator,\footnote{These are also called left-right propagators since often the ladder diagrams are drawn vertically instead of horizontally, in which case the ladder rung propagates from left to right. } which is obtained by simply continuing the Euclidean propagator $\tau \mapsto it + \frac{\beta}{2}$:
\begin{eqnarray}
G^{x}_{lr}(t) = b_x \frac{\pi^{2 \Delta_x}}{\left(\beta \cosh\left(\frac{\pi t}{\beta}\right)\right)^{2\Delta_x}}\ .
\end{eqnarray}

Here $x$ denotes $\psi$ or $\phi$ and $b_x$ denotes $A$ or $B$ respectively. The form of this propagator is the same for fermions and scalars since we only need to consider $\tau>0$ here. \\
We can then write down the expressions for the kernels. Note that each vertex gets a factor $i$ from inserting it on a Lorentzian time fold in the contour and apart from this we also give $K_{11}$ and $K_{21}$ an additional minus sign due to the ordering of the contour (see also \citep{Peng:2017kro, Peng:2017spg}). The resulting form of the kernels is:\footnote{We thank Pengfei Zhang for pointing out an error in the kernels that, in a previous version of this work, resulted in a $M/N$ dependent Lyapunov exponent.}
\begin{eqnarray}\label{kernels}
\notag K_{11} &=& 2 \, \sqrt{\frac{M}{N}} \,J \, G_R^{\psi} (t_{13}) \,G_R^{\psi} (t_{24}) \, G_{lr}^{\phi} (t_{34})\ , \\
K_{12} &=& -2 \,\sqrt{\frac{M}{N}} J \, G_R^{\psi} (t_{13}) \,  G_R^{\psi} (t_{24}) \, G_{lr}^{\psi} (t_{34})\ ,  \\
\notag K_{21} &=& 2 \,\sqrt{\frac{N}{M}}\, J \, G_R^{\phi} (t_{13}) \,  G_R^{\phi} (t_{24}) \, G_{lr}^{\psi} (t_{34})\ .
\end{eqnarray}

Where the times $t_{ij} =t_i -t_j $ are shown in \autoref{kernelfigs}.

\subsection{Integral matrix equation}
\begin{figure}
\captionsetup{singlelinecheck = false, format= hang, justification=raggedright, font=footnotesize, labelsep=space}
\includegraphics[width=\textwidth]{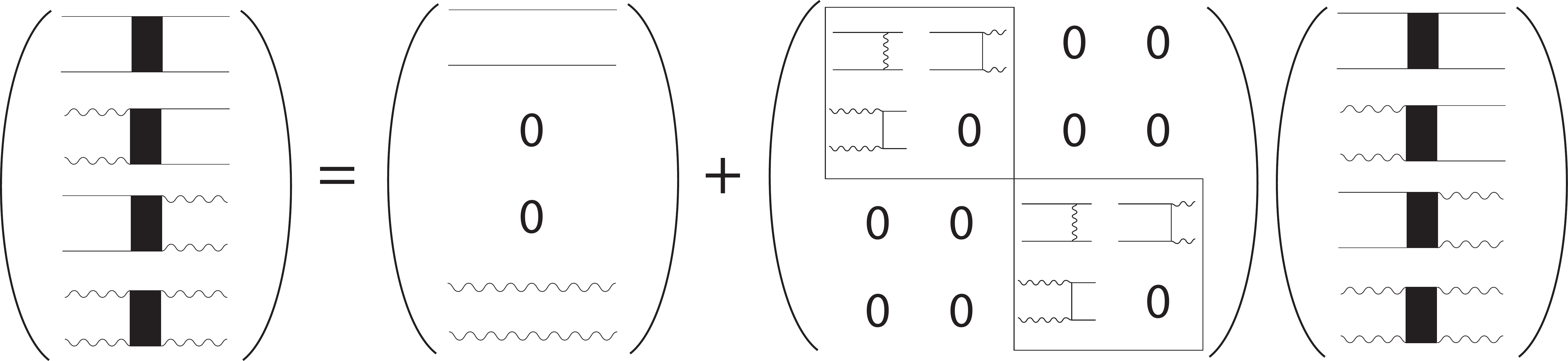}
\caption{Here we show the matrix integral equation that the OTOCs obey. The black boxes indicate the arbitrary large amount of rungs in the ladder diagrams. The very left vector consists out of all the OTOCs, the first vector on the right hand side denotes the zeroth order contributions to these and the last term is the kernel acting upon the vector of the four point functions. The matrix product also includes a convolution between the kernels and the OTOCs. Notice that the $4\times 4$ kernel matrix has a $2\times 2$ block diagonal structure.}
\label{kernelcalc}
\end{figure}

Now that we have obtained the retarded kernels we go back to our four out of time order correlators. All together they obey an integral matrix equation as shown in \autoref{kernelcalc}, this is a generalization of the one particle version seen for example in \cite{Murugan:2017eto}. In the figure we have put all the OTOCs in a four component vector seen on the very left (and right) side. These are exactly the OTOCs we named $F_{\psi \psi}$, $F_{\psi \phi}$, $F_{\phi \psi}$ and $F_{\phi \phi}$ before. Our (drawing) conventions are such that for the very left vector the times $t_1$ and $t_2$ are on the top left and bottom left of each four point function, respectively. \\

The first vector on the right hand side denotes the free contributions to the four point functions. Clearly $F_{\psi \phi}$ and $F_{\phi \psi}$ don't have these since there is no such free propagator. The matrix consists out of the retarded kernels discussed above and depicted in \autoref{kernelfigs}. Note that the matrix product in the last term also has an implicit convolution (which we will explicitly compute later on). \\

Quantum chaos implies that for late enough times these OTOCs will show exponentially growing behaviour, as discussed shortly in the introduction of this section. So let us make the following exponential growth ansatz:
\begin{eqnarray}\label{expans}
F_{ij}(t_1,t_2) = f_{ij}(t_1-t_2) \, e^{\frac{\lambda_L}{2}(t_1+t_2)} \ .
\end{eqnarray}

Where $i,j$ can denote $\psi$ or $\phi$ and $f_{ij}$ denote functions of the time difference. Under the assumption of exponential growth the matrix equation \autoref{kernelcalc} will simplify due to suppression of the zeroth order contributions. In fact, as one can easily check, the free diagrams will exponentially vanish compared to the exponential growth of the other terms. We are then left with the following equation:
\begin{eqnarray}\label{mateqn4}
\begin{pmatrix}
F_{\psi \psi}\\
F_{\phi \psi}\\
F_{\psi \phi}\\
F_{\phi \phi}
\end{pmatrix}
=
\begin{pmatrix}
    K_{11} & K_{12} & 0 & 0  \\
    K_{21} & 0 & 0 & 0 \\
    0 & 0 &K_{11} & K_{12} \\
    0 & 0 & K_{21} &0

\end{pmatrix}
\begin{pmatrix}
F_{\psi \psi}\\
F_{\phi \psi}\\
F_{\psi \phi}\\
F_{\phi \phi}
\end{pmatrix} \ .
\end{eqnarray}

Where the $F$s now obey the ansatz \autoref{expans} and the matrix multiplication still involves the convolutions. However, we see that the problem can in fact be reduced to two identical problems because of the block diagonal structure. Hence we don't have to refer to the outgoing lines of the OTOCs (the second subscript of the $F_{ij}$) and consider simply:
\begin{eqnarray}\label{mateqn}
\begin{pmatrix}
F_{\psi}\\
F_{\phi}
\end{pmatrix}
=
\begin{pmatrix}
    K_{11} & K_{12}  \\
    K_{21} & 0 \\

\end{pmatrix}
\begin{pmatrix}
F_{\psi}\\
F_{\phi}
\end{pmatrix} \ .
\end{eqnarray}

This leads to the following equations:
\begin{align}\label{matu}
&F_{\psi}(t_1,t_2) = \int\limits_{-\infty}^{t_1} dt_3 \, \int\limits_{-\infty}^{t_2} dt_4 \left[ K_{11}(t_1,t_2;t_3,t_4) \, F_{\psi}(t_3,t_4)+K_{12}(t_1,t_2;t_3,t_4) \, F_{\phi}(t_3,t_4) \right] \ , \\
&F_{\phi}(t_1,t_2) =  \int\limits_{-\infty}^{t_1} dt_3 \, \int\limits_{-\infty}^{t_2} dt_4 \, K_{21}(t_1,t_2;t_3,t_4) F_{\psi}(t_3,t_4) \ .
\end{align}

Where we have now explicitly written out the convolutions. The two equations are mixed and can be combined to give:
\begin{eqnarray}\label{conveqn}
F_{\psi}(t_1,t_2) = (K_{11} \ast F_{\psi})(t_1,t_2)+ (K_{12} \ast (K_{21} \ast F_{\psi}))(t_1,t_2) \ .
\end{eqnarray}
\subsection{Lyapunov exponents}
To actually solve the integrals we need to find the functions $f_i(t_{12})$ in \autoref{expans} such that \autoref{mateqn} is satisfied. We take the following form of the functions, similar to \cite{Maldacena:2016hyu,Murugan:2017eto,Peng:2017kro,Peng:2017spg}:
\begin{eqnarray}
F_{\psi}(t_1,t_2) &=& C_{\psi} \,\frac{e^{-\frac{\pi h}{\beta}(t_1+t_2)}}{\left(\frac{\beta}{\pi}\cosh\left(\frac{\pi t_{12}}{\beta}\right)\right)^{2\Delta_{\psi}-h}}\ , \\
\notag F_{\phi}(t_1,t_2) &=& C_{\phi} \,\frac{e^{-\frac{\pi h}{\beta}(t_1+t_2)}}{\left(\frac{\beta}{\pi}\cosh\left(\frac{\pi t_{12}}{\beta}\right)\right)^{2\Delta_{\phi}-h}}\ .
\end{eqnarray}

Where the $C_i$ denote non-zero real constants, and we have $h$ as the free exponential growth parameter. The Lyapunov exponent can be found by $\lambda_L=-\frac{2\pi h}{\beta}$.\\
 The crucial integral for the computations is as follows\footnote{As a side note, one could use substitutions of the form $z = e^{i\tau}$ to simplify the integrals, making it easier to solve them. This is done in e.g. \cite{Murugan:2017eto,Maldacena:2016hyu}.}
\begin{eqnarray}\label{intid}
&\notag \int\limits_{-\infty}^{t_1}dt_3 \int\limits_{-\infty}^{t_2}dt_4 \, \left( \frac{\pi}{\beta \sinh\left(\frac{\pi t_{13}}{\beta}\right)}  \right)^{\frac{2}{d}} \, \left( \frac{\pi}{\beta \sinh\left(\frac{\pi t_{24}}{\beta}\right)}  \right)^{\frac{2}{d}} \, \left( \frac{\pi}{\beta \cosh\left(\frac{\pi t_{34}}{\beta}\right)}  \right)^{2-\frac{4}{d}}  \frac{e^{-\frac{\pi h}{\beta}(t_3+t_4)}}{\left(\frac{\beta}{\pi}\cosh\left(\frac{\pi t_{34}}{\beta}\right)\right)^{\frac{2}{d}-h}} =\\
&  = \frac{\Gamma\left(\frac{d-2}{d}\right)^2 \, \Gamma\left(\frac{2}{d}-h\right)}{\Gamma\left(-h-\frac{2}{d}+2\right)} \frac{e^{-\frac{\pi h}{\beta}(t_1+t_2)}}{\left(\frac{\beta}{\pi}\cosh\left(\frac{\pi t_{12}}{\beta}\right)\right)^{\frac{2}{d}-h}}\ .
\end{eqnarray}

Using the above identity we can then calculate the following integrals, reminiscent of eigenvalue equations:
\begin{eqnarray}
\notag \int dt_3 \, dt_4 \, K_{11}(t_1,t_2;t_3,t_4) F^{\psi}(t_3,t_4) &=& k_{11} \, F^{\psi}(t_1,t_2)\ , \\
\label{writmat2}
\int dt_3 \, dt_4 \, K_{12}(t_1,t_2;t_3,t_4) F^{\phi}(t_3,t_4) &=&\frac{C_{\phi}}{C_{\psi}} \, k_{12} \, F^{\psi}(t_1,t_2)\ , \\
\notag \int dt_3 \, dt_4 \, K_{21}(t_1,t_2;t_3,t_4) F^{\psi}(t_3,t_4) &=& \frac{C_{\psi}}{C_{\phi}}\, k_{21} \, F^{\phi}(t_1,t_2) \ .
\end{eqnarray}

We use the following notation:
\begin{eqnarray}
 \Delta_{\psi} &=& \frac{1}{d} \ , \\
\Delta_{\phi} &=& 1-\frac{2}{d} \ ,
\end{eqnarray} 
Where the last line follows immediately from the relation between the conformal dimensions \autoref{constr1mn}. Then after some calculations we obtain the values of the $k_{ij}$:
\begin{align}\label{ev1}
&k_{11} = 8 \, \sqrt{\frac{M}{N}} \, J \, A^2  B \, \cos^2 \left(\frac{\pi}{d} \right)  \, \frac{\Gamma\left(\frac{d-2}{d}\right)^2 \, \Gamma\left(\frac{2}{d}-h\right)}{\Gamma\left(-h-\frac{2}{d}+2\right)} \ , \\
\label{ev2}
&k_{12} = -8 \, \sqrt{\frac{M}{N}} \, J \, A^3 \, \cos^2 \left(\frac{\pi}{d} \right)  \, \frac{\Gamma\left(\frac{d-2}{d}\right)^2 \, \Gamma\left(\frac{2}{d}-h\right)}{\Gamma\left(-h-\frac{2}{d}+2\right)} \ , \\
\label{ev3}
&k_{21} = -8 \, \sqrt{\frac{N}{M}} \,J \, A  B^2 \, \sin^2 \left(\pi \left(1-\frac{2}{d}\right) \right)  \, \frac{\Gamma\left(\frac{4}{d}-1\right)^2 \, \Gamma\left(\frac{2 \, (d-2)}{d}-h\right)}{\Gamma\left(\frac{4}{d}-h\right)} \ .
\end{align}

Note that here $A$ and $B$ are the coefficients of the (retarded) propagators, for which we only have an expression for $A^2B$, see \autoref{coneq}. We can now use the above $k_{ij}$ along with \autoref{writmat2} in the integral equation \autoref{conveqn}, to get:
\begin{eqnarray}\label{masteqn}
F_{\psi}(t_1,t_2) = (k_{11}+ k_{21} \, k_{12}) \, F_{\psi} (t_1,t_2) \ .
\end{eqnarray}

Of course, one could also have used the eigenfunctions (\autoref{writmat2}) first in \autoref{matu} and afterwards solved the mixing. Either way the equation resulting from the chaos regime is:
\begin{eqnarray}
k_{11}+ k_{21} \, k_{12} = 1 \ .
\end{eqnarray}

We pick then some fixed $M/N$ (which fixes $\Delta_{\psi}$ and $A^2B$) and numerically solve this equation for the Lyapunov exponent $\lambda_L = - \frac{2 \,\pi \, h}{\beta}$, which yields the solution $h=-1$. As it turns out, for $h=-1$ all the $M/N$ dependence drops out and we find in fact maximal chaos for all values of $M/N$:
\begin{eqnarray}
\lambda_L = \frac{2 \,\pi }{\beta} \ .
\end{eqnarray}

Motivated by these numerical results we analytically checked whether $k_{11}+ k_{21} \, k_{12} = 1$ for $h=-1$. To do so we use the identities from \autoref{gamid} and also the following:
\begin{eqnarray}
\sin(\pi z) &=& \frac{\pi}{\Gamma(z)\,\Gamma(1-z)} \ .
\end{eqnarray}

Using these identities and the expressions for $A^2B$, we obtain the following simplified expressions, valid at $h=-1$:
\begin{eqnarray}
k_{11} &=& \frac{\Delta_{\psi}}{1-\Delta_{\psi}} \ , \\
k_{12}\, k_{21} &=& 1-\frac{\Delta_{\psi}}{1-\Delta_{\psi}} \ .
\end{eqnarray}

Hence even though $k_{11}$ and $k_{12}\, k_{21}$ individually depend on $M/N$ (since $\Delta_{\psi}$ does), the combined result exactly cancels. \\

Lastly let us shortly mention another method of obtaining the chaos, outlined in \cite{Peng:2017kro, Peng:2017spg}. In these articles the approach is to take the matrix of kernel eigenvalues, the $k_{ij}$, and diagonalize it. Afterwards one of the eigenvalues is set to one. Let us consider this matrix:
\begin{eqnarray}
\begin{pmatrix}
    k_{11} & k_{12}  \\
    k_{21} & 0 \\
\end{pmatrix} \ ,
\end{eqnarray}
for which the resulting eigenvalues are $k_{\pm} = \frac{1}{2}\left( k_{11}\pm \sqrt{k_{11}^2+4\,k_{21}\,k_{12}} \right)$. The growing behaviour is found when $k_{+}=1$, which amounts to $k_{11}+k_{21}k_{12}=1$, consistent with our method.

\section{Discussion}\label{sectdis}
In this article we have investigated new SYK-like models with $M$ bosons and $N$ fermions. The parameter $M/N$  determines the behaviour of the model and for $M=N$ our model is related to the supersymmetric SYK model. We have found that there are two families of solutions in the model, distinguished by their conformal dimensions which we plotted as a function of $M/N$ in Figure 3. For $M=N$ the rational solution coincides with the supersymmetric solution found in \citep{Fu:2016vas}. We have shown that this branch is the dominant saddle for $M=N$. It would be interesting to see if this remains true for arbitrary $M/N$, it might be that at some point the two branches switch from being (sub)dominant. \\

Apart from this we investigated the Lyapunov exponent $\lambda_L = -\frac{2 \pi h}{\beta}$, which shows exactly maximal chaos independent of $M/N$. This is due to some non trivial cancellations in the $M/N$ dependences at $h=-1$. Due to the maximal chaos the model has a holographic interpretation as a black hole. It would be interesting to understand the role of $M/N$ in this holographic description. 
Concretely it would be interesting to find the Schwarzian for this model, in particular the coefficient in front of the Schwarzian action, related to the heat capacity, and its dependence on $M/N$. We leave this for future research.

\begin{acknowledgments}
It is a great pleasure to acknowledge useful discussions and correspondence with Dio Anninos, Tarek Anous, Micha Berkooz, Jan de Boer, Umut G\"{u}rsoy, Juan Maldacena, Cheng Peng and Moshe Rozali. We especially thank Pengfei Zhang for pointing out an error in a previous version. This work is supported in part by the D-ITP consortium, a program of the Netherlands Organization for Scientific Research (NWO) that is funded by the Dutch Ministry of Education, Culture and Science (OCW), and by the FOM programme "Scanning New Horizons".
\end{acknowledgments}

\appendix
\section{The model for a $q$-point interaction}\label{qapp}
In this appendix we will shortly show how the model and some results change when we consider an interaction vertex of degree $q$, so an interaction with one boson and $q-1$ fermions. The integer $q$ is supposed to be odd, but later we will continue it to arbitrary real values. The model we consider in the main text has $q=3$. When we apply this, the coupling $\frac{i}{2}C_{aij}\phi^a \psi^i \psi^j$ goes to $\frac{i}{(q-1)!}C_{ai_1i_2...i_{q-1}}\phi^a \psi^{i_1}...\psi^{i_{q-1}}$. Integrating out the bosons would lead to a Hamiltonian with a vertex containing $2q-2$ fermions. We further assume that $q\ll N$.\\

The disorder average is now chosen as follows:
\begin{eqnarray}
\langle C^2 \rangle = \frac{(q-1)! J}{N^{-3/2+q}M^{1/2}}\ .
\end{eqnarray}
Once again we can take the conformal form (see \autoref{ansz}) for the two point functions. By following the computations done for the $q=3$ case we find that the conformal symmetry is present under the condition that (compare to \autoref{constr1mn}):
\begin{eqnarray}
\Delta_{\phi}+ (q-1) \Delta_{\psi}=1\ .
\end{eqnarray}
The equations for the constants in the two point functions, $A^{q-1}B$, yield (compared to \autoref{coneq}):
\begin{eqnarray}\label{genqconst}
A^{q-1} B &=& \sqrt{\frac{N}{M}} \, \frac{(1-2 \Delta_{\psi}) \tan (\pi \Delta_{\psi})}{2 \pi (q-1) J}\ , \\
A^{q-1}B &=& \sqrt{\frac{M}{N}} \, \frac{(1-2(q-1)\Delta_{\psi})}{2 \pi J \tan(\pi(q-1)\Delta_{\psi})}\ .
\end{eqnarray}

The resulting transcendental equation for the conformal dimensions reads:
\begin{eqnarray}\label{generalqa2b}
\tan(\pi \Delta_{\psi}) \tan (\pi (q-1) \Delta_{\psi}) = (q-1) \frac{1-2(q-1)\Delta_{\psi}}{1-2 \Delta_{\psi}}\ .
\end{eqnarray}

One may check that the rational value $\Delta_{\psi}=\frac{1}{2q}$ is always a solution.

\bibliography{references}

\end{document}